\documentclass[prb,twocolumn,showpacs,superscriptaddress]{revtex4}
\usepackage{graphicx}
\begin{document}
\title{ Exact Mappings between Classical and Topological Orders in Two Dimensional Spin Models}
\author{Han-Dong Chen}
\affiliation{Department of Physics, University of Illinois at Urbana-Champaign, Urbana, IL 61801}
\author{Jiangping Hu}
\affiliation{Department of Physics, Purdue University,  West Lafayette, Indiana 47907}
\date{\today}
\begin{abstract}
Motivated by the duality between site-centered spin and bond-centered spin
in one-dimensional system, which  connects two different
constructions of fermions from the same set of Majorana fermions, we
show that  two-dimensional models with topological orders can be
constructed from certain well-known models with classical orders
characterized by symmetry-breaking. Topology-dependent ground state degeneracy, vanishing two-point correlation functions, and unpaired Majorana fermions on boundaries emerge naturally from such construction.
The approach opens a new way to construct and characterize topological orders.
\end{abstract}

\maketitle

Recently, topological orders have attracted intensive interests
for different reasons\cite{WEN1989,WEN1990,Wen1990a,Read2000,Wen2003,Kitaev2003,Kitaev2006}.
 The best studied example of topological order is the fractional quantum hall (FQH) states\cite{Laughlin1983}. All different FQH states have the same symmetry.  Unlike classically ordered state, FQH liquids cannot be described by Landau's  theory of symmetry breaking and the related order parameters\cite{WEN1990, Wen1995}. A new theory of topological order is proposed to describe FQH liquids\cite{Wen1995}. New nonlocal quantities, instead of local order parameters, such as ground state degeneracy\cite{WEN1989}, the non-Abelian Berry's phase\cite{Wen1990a} and topological entropy\cite{Kitaev2006,Levin2006}, were introduced to characterize different topological orders. Topological ordered systems have also been designed and studied in the context of quantum computation as an realization of potentially  fault-tolerant quantum memory and quantum computation\cite{Kitaev2003,Freedman2002,Kitaev2006}. It is the non-locality of the topological orders that significantly reduces the effect of de-coherence\cite{Dennis2002}.

     Theoretically,  a number of soluble or quasi-soluble models which capture the topological orders have
      been proposed and studied\cite{READ1991,Moessner2001a,Misguich2002a,Motrunich2002,Wen2003,Kitaev2003,Kitaev2006}.
      However, unlike the conventional orders which are entirely characterized by broken symmetries,
      the topological orders have not been characterized in a
       universal way. In fact, topological orders have to be studied case by case in different models.  Recently, it has also been pointed out that the spectrum is completely inconsequential to topological quantum order\cite{Nussinov2006} and hidden order parameter has been suggested in Kitaev model on honeycomb lattice\cite{Feng2007}. In this work, we show a new way to characterize topological  orders, which is based on well-known conventional models.  First, we show that it is  possible  to map a model with topological order to  a model with a local order parameter in certain physical realizations through a non-local duality transformation. A local order parameter  description of topologically ordered systems is potentially useful. For instance, thermodynamic properties and energy spectrum can be easily computed in terms of classical order parameters. Second,  we show that topologically ordered systems can be constructed or designed from well-studied classically ordered states by including  a topological boundary term associated with the lattice topology. In such a construction, topological properties are manifestly presented. The result provides  a novel approach for easier and/or better design of physical implementations of topological orders for quantum computation, starting from ordered systems well-understood in the framework of Landau's symmetry breaking theory.

\begin{figure}
\includegraphics[width=2.4in]{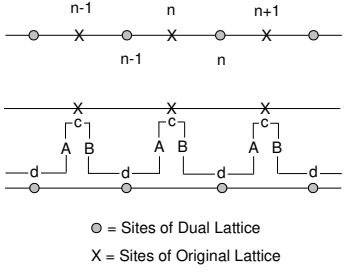}
\caption{Duality mapping in one dimension as a recombination of
Majorana fermions. $c(n)=[A(n)+iB(n)]/2$ is the fermion defined on
the original lattice site  and $d(n)=[B(n)+iA(n+1)]/2$ is defined on
the dual lattice site.}\label{FIG-duality}
\end{figure}

We start with examining a well-known nonlocal transformation,
namely, the duality between site-centered spin and bond-centered
spin in one-dimensional spin-1/2 system\cite{Fradkin1978},
\begin{eqnarray}
\mu_z(n)&=&\sigma_x(n+1)\sigma_x(n)\label{EQ-dual1}\\
\mu_x(n)&=&\prod_{m\leq n}\sigma_z(m)\label{EQ-dual2}.
\end{eqnarray}
The spin operators $\sigma$ on the original lattice can be fermionized by a Jordan-Wigner transformation
%\begin{eqnarray}
%\sigma_x(n)+i\sigma_y(n)&=&2\prod_{m<n}\sigma_z(m) c^\dag(n)\\
%\sigma_z(n)&=& 2c^\dag(n)c^{}(n)-1
%\end{eqnarray}
%where $c(n)$ and $c^\dag(n)$ are the annihilation and creation operators of the spinless fermion on lattice site $n$. We can also define Majorana fermions on site $n$
%\begin{eqnarray}
%A(n)= c^\dag(n)+c^{}(n),\quad B(n)=i\left(c^\dag(n)-c^{}(n)\right)
%\end{eqnarray}
%to rewrite the Jordan-Wigner transformation as
\begin{eqnarray}
\sigma_x(n)&=&\left[\prod_{m<n}iA(m)B(m)\right]A(n)\\
\sigma_y(n)&=&-\left[\prod_{m<n}iA(m)B(m)\right]B(n)\\
\sigma_z(n)&=&iA(n)B(n),
\end{eqnarray}
where $A(n)$ and $B(n)$ are Majorana fermions on site $n$. Fermions can be defined as $c(n)=[A(n)+iB(n)]/2$.
The duality transformation of Eq.(\ref{EQ-dual1}) and Eq.(\ref{EQ-dual2}) now reads
\begin{eqnarray}
\mu_z(n)&=&iB(n)A(n+1)\\
\mu_x(n)&=&\left[\prod_{m<n}iB(m)A(m+1)\right]B(n),
\end{eqnarray}
which is another Jordan-Wigner transformation if we
introduce a new set of fermions $d(n)=[B(n)+iA(n+1)]/2$ on the dual
lattice. It is thus clear that the duality transformation connects
two different constructions of fermions from the same set of
Majorana fermions, as illustrated in Fig.\ref{FIG-duality}.
The duality now appears as a very local transformation. However, in
terms of spin operators, it is inherently nonlocal. In the
following, we  generalize this duality transformation to
two-dimensional systems and show that the transformation can be used
to exactly map a classically ordered system to a topologically
ordered one. Two specific models are discussed.

The first model is Wen's exactly soluble spin model defined on a
square lattice\cite{Wen2003}
\begin{eqnarray}
H=g\sum_{ij}F_{ij}=g\sum_{ij} \sigma^y_{i,j}\sigma^x_{i+1,j}\sigma^y_{i+1,j+1}\sigma^x_{i,j+1},
\label{EQ-H}
\end{eqnarray}
where $(i,j)$ is the coordinate of lattice site. It is easy to see $[F_{ij},F_{i'j'}]=0$ and the
model is thus exactly soluble. This model is shown to have robust topologically degenerate ground
states and gapless edge excitations\cite{Wen2003}.

Let us first introduce the two-dimensional Jordan-Wigner transformation to fermionize the model\cite{Chen2007}
\begin{eqnarray}
\sigma^x_{ij}+i\sigma^y_{ij}&=&2\left[\prod_{j'<j}\prod_{i'}\sigma^z_{i'j'}\right]
\left[\prod_{i'<i}\sigma^z_{i'j}\right]c^\dag_{ij}\label{EQ-JW1}\\
\sigma^z_{ij}&=&2c^\dag_{ij}c^{}_{ij}-1.\label{EQ-JW2}
\end{eqnarray}
If we define the Majorana operators
\begin{eqnarray}
A_{ij}=\left(c^\dag_{ij}+ c^{}_{ij}\right)\quad \text{and}\quad
B_{ij}=i\left(c^\dag_{ij}- c^{}_{ij}\right),
\end{eqnarray}
we find $F_{ij}$
\begin{eqnarray}
F_{ij}=A_{ij}A_{i+1,j}B_{i,j+1}B_{i+1,j+1}.\label{EQ-ring}
\end{eqnarray}

\begin{figure}
\[\begin{array}{cc}
\includegraphics[width=1.5in]{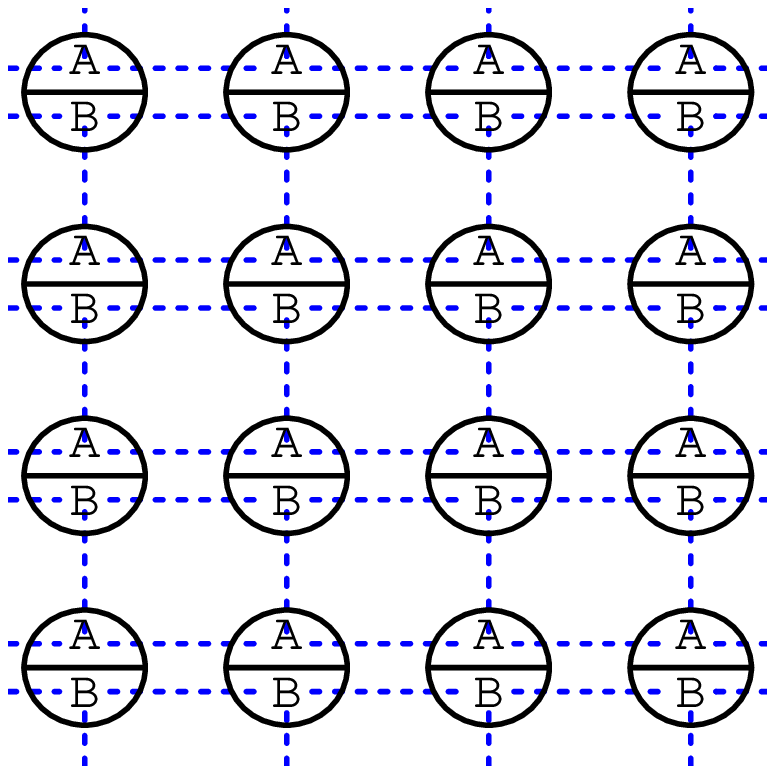}&
\includegraphics[width=1.5in]{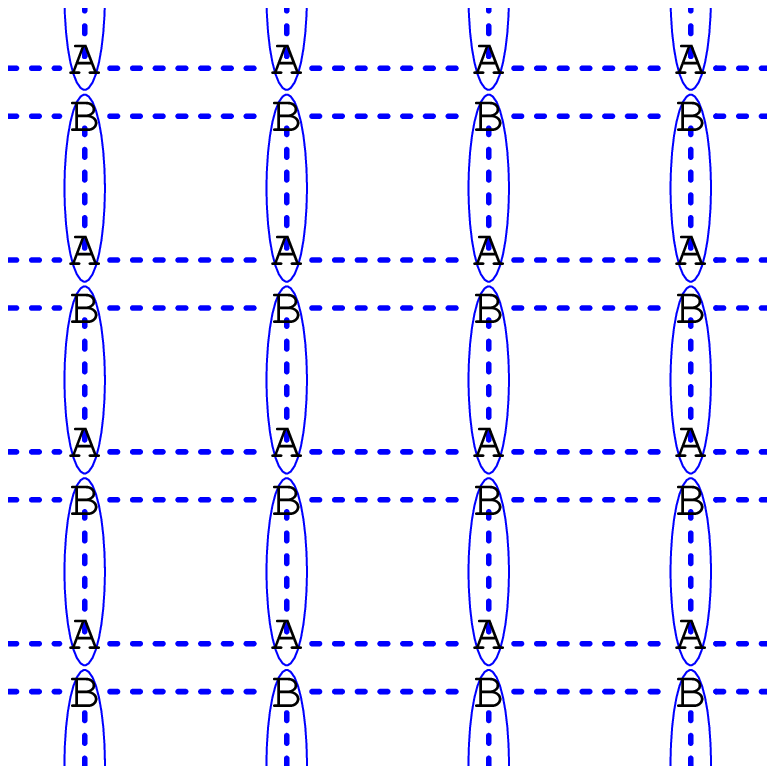}\\
(a) & (b)
\end{array}\]
\caption{(a) Graphical representation of Wen model. Each circle denotes one original lattice site on which two Majorana fermions $A$ and $B$ are defined. Each dotted rectangle contributes a ring-exchange term given by Eq.(\ref{EQ-ring}). (b) After introducing fermions on vertical bonds, the model reduces to decoupled Ising chains.}\label{FIG-ring}
\end{figure}

It is now interesting to generalize the duality to two dimensions and define
fermions on vertical bonds $(i,j)-(i,j+1)$
\begin{eqnarray}
d_{ij}=\left(A_{ij}+iB_{i,j+1}\right)/2\\
d_{ij}^\dag=\left(A_{ij}-iB_{i,j+1}\right)/2.
\end{eqnarray}
It follows that
\begin{eqnarray}
iA_{ij}B_{i,j+1}=2d^\dag_{ij}d_{ij}-1\equiv \mu_{ij}^z,
\end{eqnarray}
where $\mu_{ij}$ is related to the fermion $d$ through a Jordan-Wigner transformation. The Wen Hamiltonian thus reads
\begin{eqnarray}
H=g\sum_{ij} \mu_{i,j}^z\mu_{i+1,j}^z.\label{EQ-Ising}
\end{eqnarray}
This new Hamiltonian describes a set of decoupled Ising chains\cite{Nussinov2006a}. A first order phase transition from ferromagnetism to antiferromagnetism happens at $g=0$.
It is now straightforward to see that all two-point correlation functions are identically zero $\langle \sigma^a_1\sigma^b_2 \rangle=0$ in the ground state. This is so because $\sigma^1_1\sigma^b_2$ contains $A_{ij}$ (or $B_{i,j+1}$) that is unpaired with its partner $B_{i,j+1}$ (or $A_{ij}$) 
%they all contain $\langle \mu^x\rangle$ and/or $\langle\mu^y\rangle$ 
due to the fractionalization of $\sigma_{ij}$ into $A_{ij}$ and $B_{ij}$ and the recombination of $A_{ij}$ and $B_{i,j+1}$ into $\mu^z_{ij}$.

It is also interesting to fermionize a Zeeman term %$B\sum_{ij}\sigma_{ij}$
\begin{eqnarray}
b\sum_{ij}\sigma_{ij}^z%=iA_{ij}B_{ij}
=b\sum_{ij}\left(d_{ij}+d_{ij}^\dag\right)
\left(d_{i,j-1}-d_{i,j-1}^\dag\right).\label{Zeeman}
\end{eqnarray}
We notice that Eq.(\ref{EQ-Ising})+Eq.(\ref{Zeeman}) is the same fermioinic Hamiltonian obtained by fermionizing quantum compass model using Jordan-Wigner transformation\cite{Chen2007}. After including a Zeeman term, the Wen's soluble model is thus equivalent to the quantum compass model, which is shown to have dimensional reduction\cite{Xu2004,Nussinov2005} and a first-order phase transition at $b=g$\cite{Chen2007}.

The duality mapping can also be made explicit as follows.
Define $\mu^z_{i,j}$ on the bond $(i,j)-(i,j+1)$
\begin{eqnarray}
\mu^z_{i,j}=\sigma^y_{i,j}
\left(\prod_{i'>i}\sigma^z_{i',j}\right)
\left(\prod_{i'<i}\sigma^z_{i',j+1}\right)\sigma^y_{i,j+1}.\label{EQ-Wen-Ising-mapping}
\end{eqnarray}

\begin{figure}
\[
\begin{array}{cc}
\includegraphics[width=1.2in]{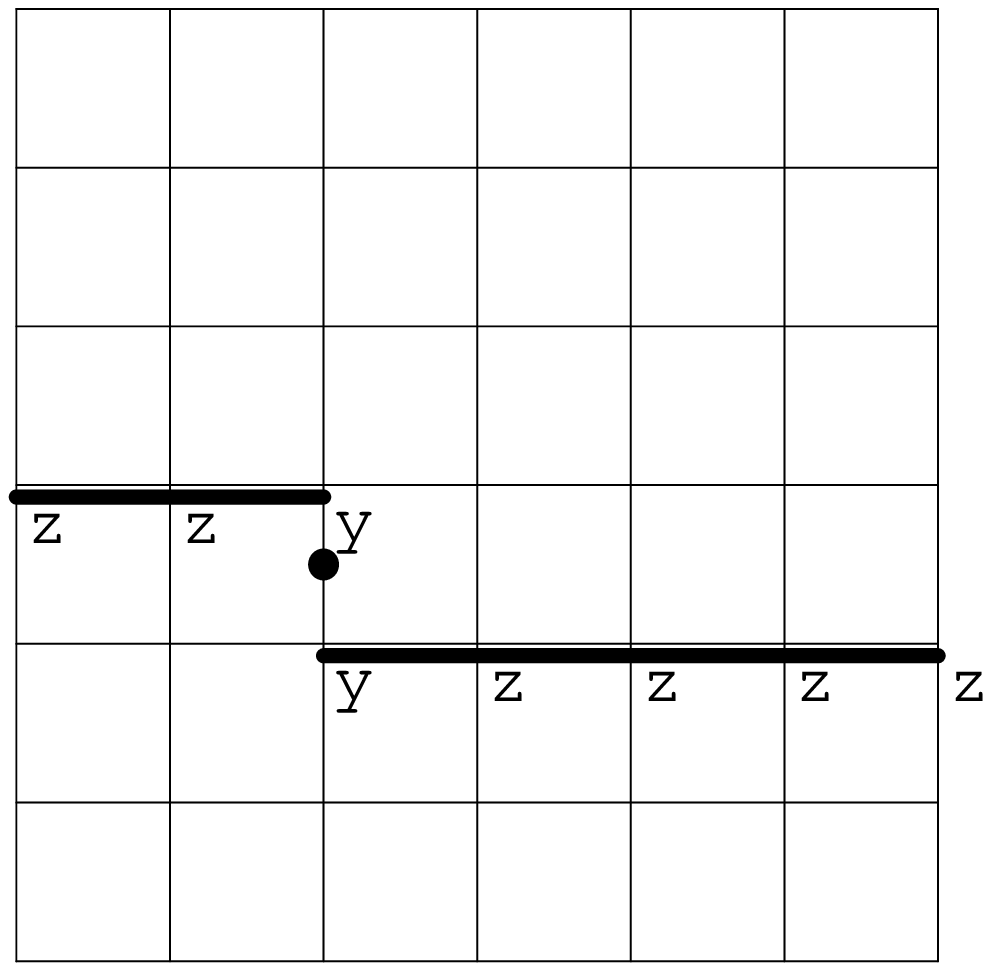}&
\includegraphics[width=1.2in]{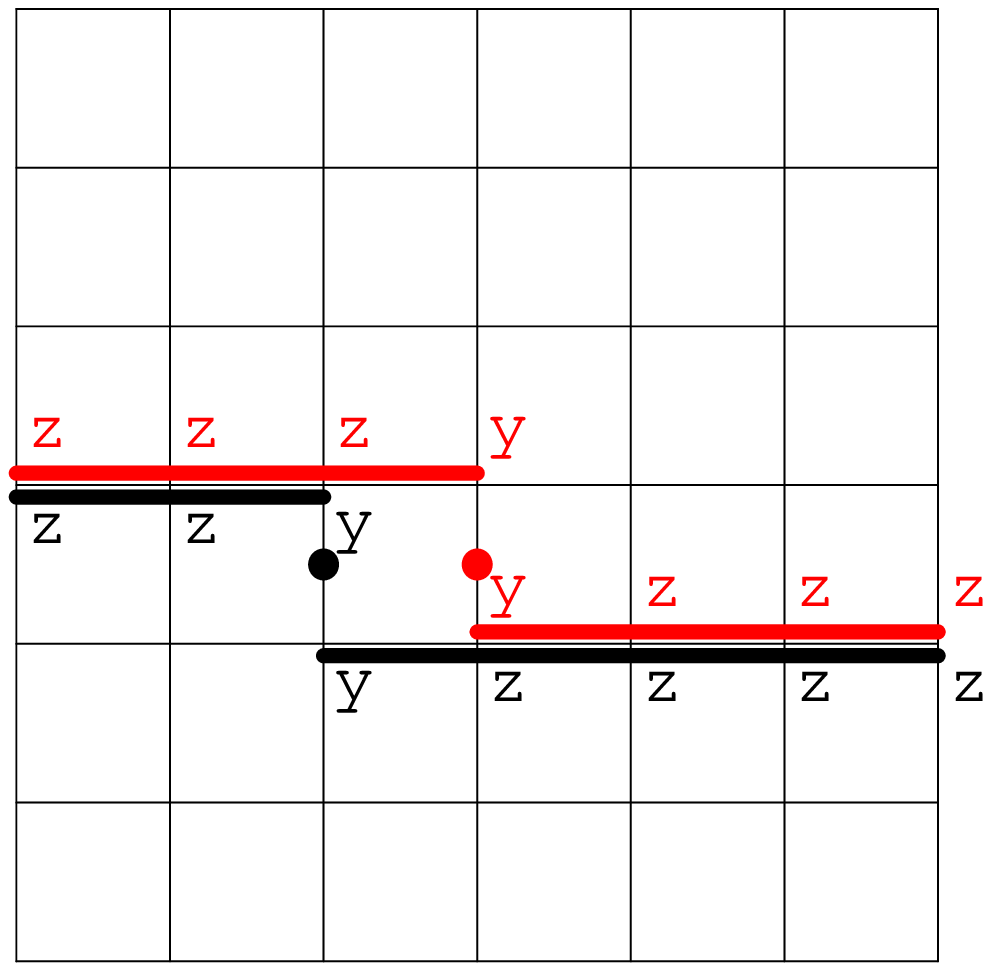}\\
(a) & (b)
\end{array}
\]
\caption{(a) Graphical representation of Eq.(\ref{EQ-Wen-Ising-mapping}).  $\mu^z$ is defined on a vertical bond as the product of spin components as labelled on the origin lattice sites along the thick dark line. (b) Nearest neighbor coupling between $\mu^z_{ij}$ and $\mu^z_{i+1,j}$.}\label{FIG-m}
\end{figure}

\noindent Let us first prove that $\mu^z_{i,j}$ commutes with $\mu^z_{k,l}$.
Without losing generality, let us assume $l\geq j$. Let us consider the overlaps between the original lattice sites involved in $\mu^z_{ij}$ and $\mu^z_{kl}$. (a) If $l-j\geq 2$, there is no overlap. Similarly, no overlap happens when $l=j+1$ and $k>i$ and two $\mu^z$ commute. (b) $l=j+1$ and $k=i$, there is only one common site on which $\sigma^y$ is involved in both $\mu^z$. (c) $l=j+1$ and $k<i$ or $l=j$ (see Fig.\ref{FIG-m}(b), as far as a commutation relation is concerned, the only relevant part is
\begin{eqnarray}
\left(\sigma^z_r\sigma^y_k\right)_1\left(\sigma^y_r\sigma_k^z\right)_2
=\left(\sigma^y_k\sigma^z_r\right)_1\left(\sigma^z_k\sigma_r^y\right)_2.
\end{eqnarray}
Here, the subscripts $1$ and $2$ denote two different sites of original square lattice while $r$(red) and $k$(black) are used to keep track of which $\mu^z$ the spin operators $\sigma$ come from. Based on the above consideration, we conclude that $\mu$'s are commuting with each other.
It is also trivial to see  $\left(\mu^z\right)^\dag=\mu^z$ and $\left(\mu^z\right)^2=1$. Therefore, $\mu^z$ can be
  viewed as an Ising degree of freedom defined on a vertical bond. Let us now consider the  interaction term between two nearest-neighboring $\mu^z$, as illustrated in Fig.\ref{FIG-m}(b),
\begin{eqnarray}
\mu^z_{i,j}\mu^z_{i+1,j}&=& \sigma^y_{i,j}\left(\sigma^z\sigma^y\right)_{i+1,j}
\left(\sigma^y\sigma^z\right)_{i+1,j}\sigma^y_{i+1,j+1}\nonumber\\
&=&\sigma^y_{i,j}\sigma^x_{i+1,j}\sigma^x_{i+1,j}\sigma^y_{i+1,j+1}.
\end{eqnarray}
Therefore, the original Hamiltonian (\ref{EQ-H}) is equivalent to Eq.(\ref{EQ-Ising}) and the
Ising order is mapped to the quantum order studied by Wen\cite{Wen2003}. The explicit mapping (\ref{EQ-Wen-Ising-mapping}) also allows us to examine the topological nature of the ordering. In the basis of $\sigma^y$, $\mu^z_{ij}$ creates two kinks at the two ends $(i,j)$ and $(i,j+1)$ of the vertical bond. Therefore, the ordering is actually a condensation of kink-dipoles. The topological nature of the order in this model is thus similar to the one of Ising lattice gauge theory\cite{Kogut1979}.
 
The discussion of the exact mapping is so far limited to bulk terms. It is important to study the boundary conditions and demonstrate explicitly the dependence of ground degeneracy on the topology. In general, the boundary conditions will induce a nonlocal phase factor to the coupling strength between boundary spins, due to the phase term in the Jordan-Wigner transformation. The topology-dependent phase factor has profound consequence as we shall show shortly. One immediate consequence is that it determines the ground state degeneracy.

The most interesting boundary condition is the case where we put the original spin model into a closed topology.
A simple closed manifold is a torus by taking periodic boundary conditions along both directions. The boundary term along $y$-direction is $H_b^y=g\sigma^y_{i,L_y}\sigma^x_{i+1,L_y}\sigma^y_{i+1,1}\sigma^x_{i,1}$. It is clear that the phase term cancels and the periodic condition along the $y$-direction is mapped to a periodic boundary condition in the direction perpendicular to the Ising chain. The periodic boundary condition along $x$-direction induces a boundary term  $H_b^x=g\sigma^y_{L_x,j}\sigma^x_{1,j}\sigma^y_{1,j+1}\sigma^x_{L_x,j+1}$. This term is mapped to $H_b^x=g_{x} \sum_j \mu_{1,j}^z\mu_{L_x,j}^z$ with the coupling strength $g_{x}$ given by
\begin{eqnarray}
g_{x}= g\prod_i{\sigma_{i,j}^z\sigma^z_{i,j+1}}
=g\prod_{i,j}iB_{ij}\mu^z_{ij}A_{i,j+1}.
\end{eqnarray}
The boundary term couples nearest neighboring chains nonlocally,
which manifestly represents the hidden topological structure in the
original model.  A direct consequence of this coupling is the
partial lift of ground degeneracy.

Another interesting case is the ribbon structure, where periodic boundary condition in $y$ direction
and open boundary condition in $x$ direction are assumed.  The open boundary condition in $x$ direction
is now mapped to the open boundary conditions in the spin chains.
Consequently, the ground state now has an effect of dimension reduction and huge degeneracy $2^{L_y}$,
 where we denote $L_x$ and $L_y$ as the system sizes along $x$ and $y$ directions respectively.
 The degeneracy can actually be related to free Majorana fermions on boundaries.
  This can be shown by considering an equivalent geometry where we set periodic boundary  condition along $x$ and open boundary condition along $y$. The ground state degeneracy is $2^{L_x}$ in this case. The mapping thus leads to unpaired Majorana fermions,  $A_{i,L_y}(i=1,..,L_x)$ on the sites of top boundary and $B_{i,1}(i=1,..,L_x)$ on the sites of bottom boundary. $A_{i,L_y}$ are coupled to the bulk system through the boundary term in the form of $\prod_{i=1}^{L_x}A_{i,L_y}$. Similarly, $B_{i,1}$ are coupled to bulk      through $\prod_{i=1}^{L_x}B_{i,1}$. Therefore, the operators that flip even umbers of unpaired Majorana fermions on top/bottom boundaries are conserved quantities and consequently lead to degenerated ground states. For instance, we can combine $A_{1,L}$ and $A_{2,L}$ into a fermion whose particle number $(iA_{1,L}A_{2,L}+1)/2$ is a conserved quantity. We thus have successfully mapped the global (nonlocal) $Z_2$ degree of freedom of the decoupled Ising chains into a local degree of freedom of unpaired Majorana fermions at the ends of the chains.

To illustrate our approach further, we show that  similar physics
can also be reached for the second Kitaev model  defined on a
honeycomb lattice\cite{Kitaev2006},
\begin{eqnarray}
H&=&-\sum_{\lambda=x,y,z}\quad\sum_{\lambda-bonds} J_\lambda
S^\lambda_{j}S^\lambda_{k}
%-J_y\sum_{y-bonds} S^y_{j}S^y_{k}\nonumber\\
%&&-J_z\sum_{z-bonds} S^z_{j}S^z_{k}
.
\end{eqnarray}
This topologically ordered model
 can be mapped to a model of spinless fermions whose ground states are characterized by local order parameters.

\begin{figure}
\includegraphics[width=2in]{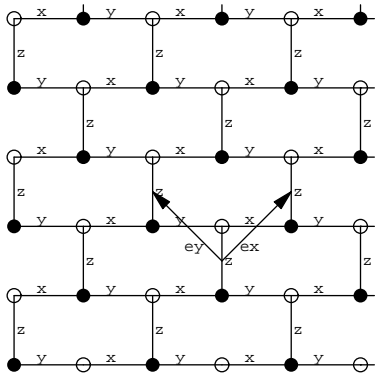}
\caption{Deformed honeycomb lattice and three types of bonds.}
\label{FIG-honeycomb}
\end{figure}

Again, we  fermionize this model using the Jordan-Wigner
transformation. The idea is to deform the honeycomb lattice into a
brickwall lattice as shown in Fig.\ref{FIG-honeycomb}.
We introduce the subscripts $b$ and $w$ to denote the white and black
sites of a bond as illustrated in Fig.\ref{FIG-honeycomb}. We also define the
corresponding Majorana fermions $ A_w=(c-c^\dag)_w/i$ and
$B_w=(c+c^\dag)_w $ for white sites, $ B_b=(c-c^\dag)_b/i$ and
$A_b=(c+c^\dag)_b $ for black sites.
After a Jordan-Wigner transformation given by
Eq.(\ref{EQ-JW1}) and Eq.(\ref{EQ-JW2}), the Hamiltonian of Kitaev model becomes
\begin{eqnarray}
H&=&-\frac{i}{4}\left[\sum_{x-bonds} J_x A_w^{}A_b^{} - \sum_{y-bonds}J_y A_b^{}A_w^{}\right]
\nonumber\\
&&-\frac{i}{4}\sum_{z-bonds} J_z \alpha_{bw}^{} A_b A_w,
\end{eqnarray}
where $\alpha=iB_{b} B_{w}$ defined on each vertical bond. It is easy to see that $\alpha$ is a conserved quantity\cite{Feng2007} and can now be taken as a number that can take either $+1$ or $-1$. The Hamiltonian is now quadratic in $A$ and readily to be solved exactly.

We are now ready to generalize the duality to brickwall lattice and
introduce fermion on a $z$-bond, $ d=(A_w+i A_b)/2$ and $d^\dag =
(A_w-iA_b)/2 $, where $A_w$ and $A_b$ are the Majorana fermions on
the white and black sites of a given $z$-bond. We thus have a model
for fermions on a square lattice with site-dependent chemical
potential
\begin{eqnarray}
4H &=&  J_x\sum_{i}\left(d_i^\dag + d_i^{}\right)
\left(d_{i+\hat{e}_x}^\dag - d_{i+\hat{e}_x}^{}\right)\nonumber\\
&&+J_y\sum_{i}\left(d_i^\dag + d_i^{}\right)
\left(d_{i+\hat{e}_y}^\dag - d_{i+\hat{e}_y}^{}\right)
\nonumber\\&&+J_z\sum_i \alpha_i (2d^\dag_i d^{}_i-1).\label{EQ-fermion-model}
\end{eqnarray}
Here $\hat{e}_y$ connects two $z$-bonds and crosses a $y$-bond, similarly for $\hat{e}_x$, as illustrated in Fig.\ref{FIG-honeycomb}.
This Hamiltonian describes a system of spinless fermions with $p$-wave BCS pairing and site-dependent
 chemical potential, where the ground states are characterized by local order parameters. Previous discussions about ground state degeneracy and vanishing two-point spin correlation functions can now be extended to this model straightforwardly. Unpaired free Majorana fermions also emerge naturally at open boundaries. For instance, a ribbon geometry can be achieved by breaking a row of $z$-bonds. For each broken $z$-bond, the $Z_2$ degree of freedom $\alpha=iB_bB_w$ is fractionalized into two unpaired free Majorana fermions, $B_b=(c-c^\dag)_b/i$ on the top boundary and $B_w=(c+c^\dag)_w$ on the bottom boundary. A detailed study of the fermionized Hamiltonian (\ref{EQ-fermion-model}) will be presented elsewhere\cite{Chen2007a}.

In summary, we have successfully constructed exact mappings from
topological orders to classical orders in two exactly soluble spin
models. The topological dependence in the
later model is manifestly  represented in the terms
resulted from the mapping of boundary conditions.
Unpaired Majorana fermions on open boundaries and vanishing two-point spin correlation functions also follow naturally from our construction.
Our work suggests a novel approach to construct certain topological orders from well-studied classical orders through a nonlocal transformation. Finally, we would like to point out that the transformation used in this work only works for a limited class of models. However, we conjecture that a general connection between topological orders and classical orders might be possible and more beautiful nonlocal transformations are waiting to be discovered.

We would like to acknowledge useful discussions with Z. Nussinov and C. Xu. H.D. Chen is supported by the U.S. Department of Energy, Division of Materials
Sciences under Award No. DEFG02-91ER45439, through the Frederick
Seitz Materials Research Laboratory at the University of Illinois at
Urbana-Champaign. J.P. Hu is supported by the National Science
Foundation under grant number: PHY-0603759.
%\bibliography{majorana,extra}

\end{document}